\documentclass[prl,twocolumn,showpacs,preprintnumbers,amsmath,amssymb]{revtex4-1}
\usepackage{amsmath}
\usepackage{graphicx}
\usepackage{dcolumn}
\usepackage{bm}
\usepackage{natbib}
\usepackage[colorlinks=true,citecolor=blue,linkcolor=blue,urlcolor=blue]{hyperref}

\newcommand{\nd}{\ensuremath{\mathrm{NdFe_3(BO_3)_4}}}

\newcommand{\SNd}{$\langle\mathrm{S_{Nd}}\rangle$}
\newcommand{\SFe}{$\langle\mathrm{S_{Fe}}\rangle$}
\newcommand{\SNdsq}{$\langle\mathrm{S_{Nd}}\rangle^2$}
\newcommand{\SFesq}{$\langle\mathrm{S_{Fe}}\rangle^2$}
\newcommand{\Bfield}{$B$}
\newcommand{\Bvec}{$\mathbf{B}$}

\begin{document}

\preprint{CAPS/123-QED}

\title{Magnetic frustration, phase competition and the magneto-electric effect in $\mathbf{NdFe_3(BO_3)_4}$}

\author{J.E. Hamann-Borrero$^1$}
  \email{j.e.hamann.borrero@ifw-dresden.de}	
\author{S. Partzsch$^1$}
\author{S. Valencia$^{2}$}
\author{C. Mazzoli$^{3,4}$}
\author{J. Herrero-Martin$^{3,5}$}
\author{R. Feyerherm$^2$}
\author{E. Dudzik$^2$}
\author{C. Hess$^1$}
\author{A. Vasiliev$^6$}
\author{L. Bezmaternykh$^7$}
\author{B. B\"{u}chner$^{1,8}$}
\author{J. Geck$^1$}
  \email{j.geck@ifw-dresden.de}

\affiliation{$^1$Leibniz Institute for Solid State and Materials Research, IFW Dresden, 01171 Dresden, Germany}
\affiliation{$^2$Helmholtz Zentrum Berlin. Albert Einstein Str.15 12489 Berlin, Germany}
\affiliation{$^3$European Synchrotron Radiation Facility (ESRF), BP 220, 38043 Grenoble, France}
\affiliation{$^4$Physics Department, Milan Politechnical University, Piazza Leonardo da Vinci 32, 20133 Milano, Italy}
\affiliation{$^5$Instituto de Ciencia de Materiales de Barcelona, CSIC. Campus de la UAB 08193, Bellaterra, Spain}
\affiliation{$^6$Low Temperature Physics department, Faculty of Physics, Moscow State University, Moscow, 119992 Russia}
\affiliation{$^7$L.V Kirensky Institute of Physics, Siberian Division, Russian Academy of Sciences, Krasnoyarsk, 660,0,36 Russia}
\affiliation{$^8$Institute for Solid State Physics, Dresden Technical University, TU-Dresden, 01062 Dresden, Germany}

\date{\today}

\begin{abstract}
We present an element selective resonant magnetic x-ray scattering study of \nd\/ as a function of temperature and applied magnetic field. Our measurements  show that the magnetic order of the Nd sublattice is induced by the Fe spin order. 
When a magnetic field is applied parallel to the hexagonal basal plane, the helicoidal spin order is suppressed and a collinear ordering, where the moments are forced to align in a direction perpendicular to the applied magnetic field, is stabilized. This result excludes a non-collinear spin order as the origin of the magnetically induced electric polarization in this compound. Instead our data imply that magnetic frustration results in a phase competition, which is the origin of the magneto-electric response.
\end{abstract}
\pacs{Valid PACS appear here}

\maketitle

\section {Introduction}

The coupling between magnetism and electric polarization in so-called multiferroic materials is a major topic of current condensed matter research, since it is of large interest for both basic science and technological applications~\cite{cheong-nature,Khomskii-phys}.  Unfortunately ferroelectric and magnetic order rarely coexist and even if they do, the coupling between them is usually very weak~\cite{Khomskii-phys}.

It was therefore greeted with great excitement when frustrated magnetic materials were discovered, where the coupling of magnetic and ferroelectric orders is extraordinarily strong. Typically these systems relieve the magnetic frustration to some extend via e.g. lattice distortions, which, as a byproduct, also create a ferroelectric polarization~\cite{cheong-nature,Khomskii-phys,kimura-nature}. In this way the magnetic and ferroelectic (FE) orders are directly connected to one another, resulting in a very strong magneto-electric coupling. 

Typical examples are non-collinear magnets~\cite{mostovoy06}, where the spin order rotates about an axis \textbf{e} and propagates along a given direction \textbf{q}. In these magnetic phases a FE polarization \textbf{P} can exist, which, according to the conventional theory, is given by $\textbf{P} \propto \textbf{e}\times \textbf{q}$~\cite{mostovoy06,ADMA200901961}. However, FE order can also be driven by collinear spin order \cite{choi:047601,cheong-nature}. In order to reveal the microscopic origin of the ME coupling it is therefore very important to know the type of magnetic order.

\begin{figure*}
\begin{minipage}{5cm}
\includegraphics[width=\columnwidth]{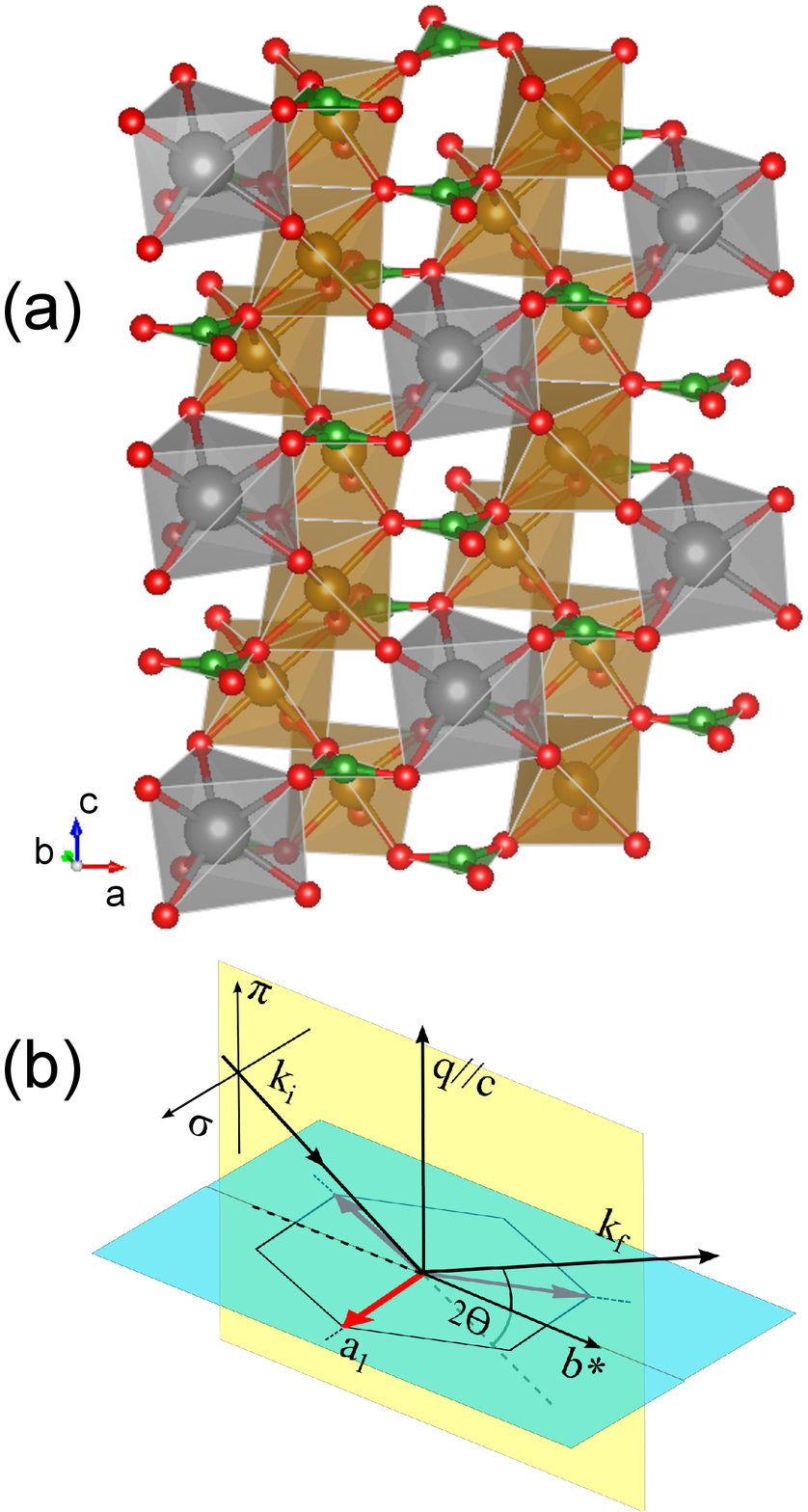}
\end{minipage}
\begin{minipage}{11cm}
 \includegraphics[angle=-90,width=\columnwidth]{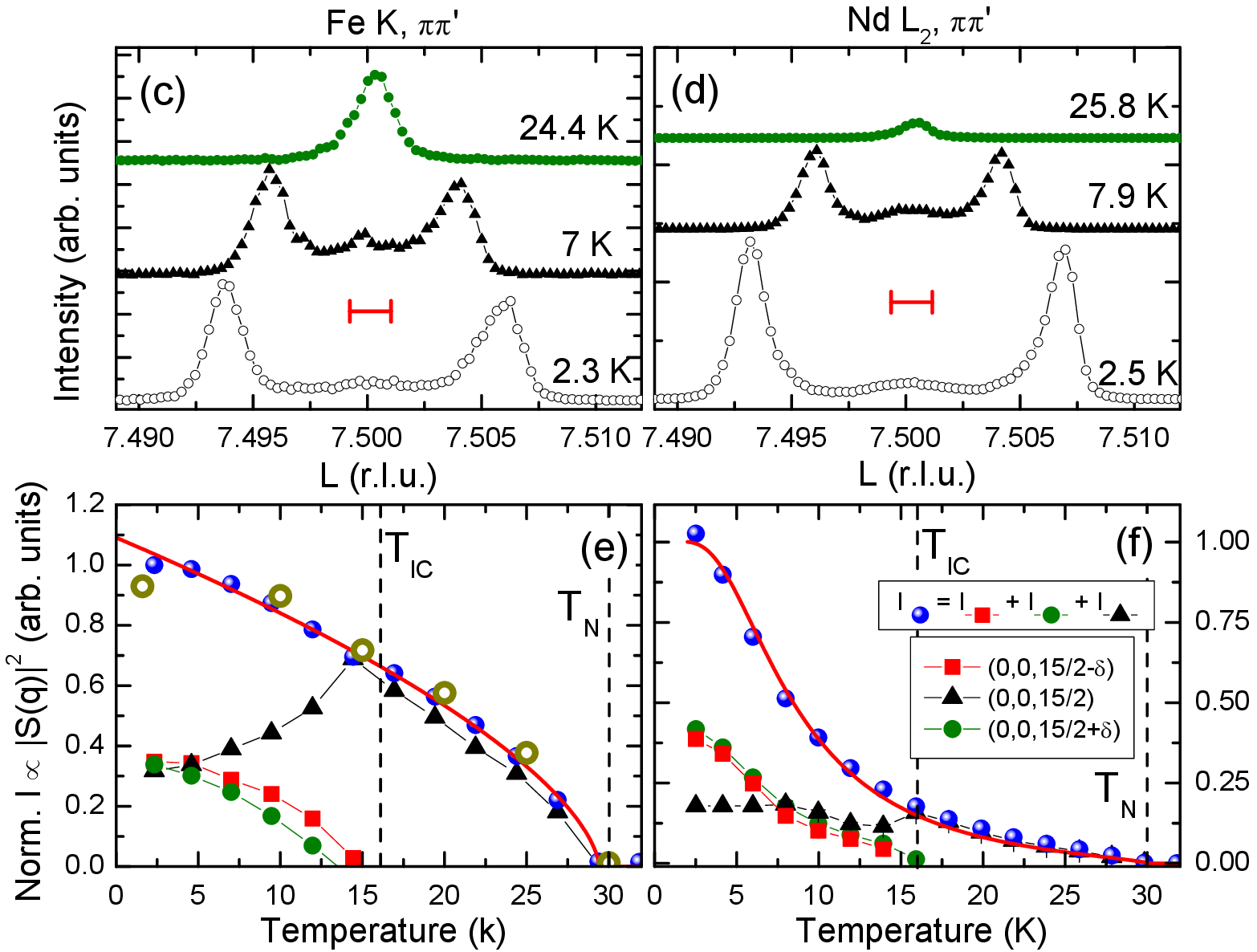}
\end{minipage}
\caption{(Color online) (a) Crystal structure of \nd~ with Fe chains of octahedra (brown), Nd prisms (gray) and the B triangles (green). (b) Experimental scattering geometry showing the scattering plane (yellow) and the hexagonal basal plane (blue) where the $a$-axis is perpendicular to the scattering plane. Temperature dependent intensities of the magnetic (0,0,15/2) reflection at the Fe K (c) and Nd L$_2$ (d) edges. The corresponding normalized integrated intensities are given in panels (e) and (f). 
Open circles in figure (e) show the T-dependent Fe moment, squared and scaled for comparison, from neutron diffraction (taken from \cite{fischer06}). Red scale in figures (c) and (d) show the (0,0,6) Bragg peak FWHM=0.0018(3)\,r.l.u. Red lines are fits to the data as explained in the text.} 
\label{fig:Tdep152-fits}
\end{figure*}

Among multiferroic systems that show non-collinear spin order, the magneto-electric (ME) \nd~ has attracted considerable attention due to the large coupling between magnetic and electric orders~\cite{kadomtseva06,kadomtseva07}. Upon cooling in zero magnetic field, a collinear magnetic order sets in first at $T_{N}\approx30$\,K, consisting of ferromagnetic hexagonal $ab$-planes, which are stacked antiferromagnetically along the perpendicular $c$-direction (cf. Fig.\,\ref{fig:Tdep152-fits})~\cite{fischer06,janoschek}. Recent polarized neutron scattering experiments~\cite{janoschek} have shown that a commensurate (C) to incommensurate (IC) transition is present at $T_{IC}\approx13.5$~K. At this temperature the magnetic structure turns into an IC spin helix that propagates along the $c$ axis. 
Although this neutron study revealed important details about the helical spin order in \nd , the microscopic origin of the spin helix remains to be clarified.

Even more interesting, at low temperatures, the electric polarization strongly increases when a magnetic field is applied parallel to $a$, reaching a maximum of $P_a\sim390~\mu\rm C/m^2$ at $B_a\sim1.3$~T and 4.5~K\cite{kadomtseva06,Kadomtseva-LTP2010} (in TbMnO$_3$ $P_c\sim800~\mu\rm C/m^2$\cite{kimura-nature}). The spin helix in zero field cannot explain the electric polarization since $\textbf{e}\parallel \textbf{q}$, for which the conventional theory yields $\textbf{P}=0$. This already implies that the spin order in applied magnetic fields must change. As stated already above, it is essential to determine these changes in order to uncover the driving force behind the ME coupling. Up to date, however, the spin order in applied magnetic fields was unknown for this compound. 

In this letter we clarify the two central questions mentioned above, namely the origin of the spin helix as well as the magnetic order in applied magnetic fields.  By means of resonant magnetic x-ray scattering (RMXS) we show that the magnetic order of the Nd-sublattice is induced by the Fe-spin order and determine the magnetic field induced changes in the spin order. The obtained data provide clear experimental evidence that the C-IC transition and the ME effect are both related to the magnetic frustration between the Nd and the Fe sublattice. 

RMXS measurements at the Nd L$_2$ (6.726~keV) and Fe K (7.128~keV) edges were performed at beamlines ID20 (ESRF in Grenoble, France~\cite{id20})  and MagS (BESSY, Berlin~\cite{Dudzik}). 
A scheme of the scattering geometry used for the present experiments is illustrated in Fig.\ref{fig:Tdep152-fits}(b). The sample was mounted inside a cryomagnet with the horizontal scattering plane parallel to the $b^*c$ plane of the sample. Measurements  down to 2~K and with magnetic fields up to 2~T along the crystallographic $a$ axis were performed, i.e., the magnetic field was aligned perpendicular to the scattering plane.
During the measurements both the incoming and the outgoing photon polarizations were controlled, where $\sigma$ ($\sigma'$) and $\pi$ ($\pi'$) refer to linearly incoming (outgoing) polarized light perpendicular and parallel to the scattering plane, respectively.  

\nd~ crystallizes in a hexagonal structure with space group $R32$ with the hexagonal unit cell parameters $a \approx 9.59\,\text{\AA}$, $c \approx 7.61\,\text{\AA}$ \cite{fischer06}. As shown in Fig.\ref{fig:Tdep152-fits}(a), the structure consists of chains of edge shared FeO$_6$ octahedra along the crystallographic $c$ axis. These chains are connected by BO$_3$ triangles and NdO$_6$ prisms. \nd~therefore hosts two coupled magnetic sublattices of, respectively, Fe  and Nd sites.

Below T$_N\sim30$~K, the appearance of superlattice reflections at $\textbf{q}=(0,0,3l\pm\frac{3}{2})$ (where $l$ is an integer) reveals a doubling of the unit cell along $c$ (see Fig. \ref{fig:Tdep152-fits}(c) and \ref{fig:Tdep152-fits}(d)). This observation is in perfect agreement with recent neutron\cite{fischer06,janoschek} as well as x-ray scattering\cite{hamann-boratesNRXS} reports. The simultaneous appearance of these reflections together with the magnetic order and the fact that the reflected photons for incoming $\pi$ polarization are mostly $\sigma$ polarized, clearly identifies these reflections as magnetic superlattice peaks.

Figs.\ref{fig:Tdep152-fits}c  and \ref{fig:Tdep152-fits}d show high-resolution scans of the (0,0,15/2) magnetic reflection at different temperatures, which were measured at the Fe K and Nd L$_2$ edge, respectively for B=0\,T. Below T$_N\approx30$\,K  a single C-reflection (green circles) is observed, corresponding to the collinear AFM structure~\cite{fischer06,janoschek,hamann-boratesNRXS}. Upon further cooling down to T$_{IC}\approx16$\,K the peak splits into two IC peaks at (0,0,15/2$\pm\delta$)  due to the formation of a long period spin helix (black triangles) that propagates along the $c$ direction\cite{fischer06,janoschek}.  Surprisingly, although the C reflection becomes very weak and broad, it never vanishes completely which indicates a coexistence of the C and IC phases at T~$<$~T$_{IC}$. 

The value of $\delta$ strongly increases below T$_{IC}$ as does the intensity of the IC reflections (open circles). 
This temperature dependent $\delta$  reveals a continuous change of the helix periodicity. Specifically, we find that, in real space, the period of the spin helix decreases from $\sim$3900~\AA{} ($\sim$523 unit cells) at 14~K to $\sim$1123~\AA{} ($\sim$146 unit cells) at 2~K. Again, these observations are in very good agreement with previous neutron results\cite{janoschek}.

Within the kinematical theory of magnetic diffraction, the RMXS integrated intensity at the Fe K and the Nd L$_2$ edge is directly related to the ordered magnetic moment on the Fe and the Nd sublattice, respectively\,\cite{blume88,gibbs1988}.
In other words, RMXS enables to study the magnetic order on the Fe and Nd sublattices separately by tuning the photon energy to the corresponding absorption edge.

Fig. \ref{fig:Tdep152-fits}(e) shows the temperature dependent integrated RMXS intensities measured at the Fe K edge and normalized to its value at the lowest temperature.
As can be observed in this figure, the C peak first increases rapidly with cooling below T$_N$. Upon further cooling below T$_{IC}$, the C reflection collapses whereas the IC reflections appear and increase in intensity.

A similar peak splitting behavior is also observed at the Nd L$_2$ edge. However, here the temperature evolution of the intensities is very different. As shown in Fig.\,\ref{fig:Tdep152-fits}(f), the intensity enhancement of the C peak is much weaker in the range T$_{IC}<$~T~$<$T$_N$. Below T$_{IC}$ the C peak intensity barely changes, while the IC peaks intensities increase pronouncedly.  

The blue circles in Figs.\,\ref{fig:Tdep152-fits}e and \ref{fig:Tdep152-fits}f correspond to the integrated intensity of the (0,0,15/2) peak in the C phase plus the summed integrated intensities of the two satellites at  (0,0,15/2$\pm\delta$) . This total integrated intensity is proportional to \SFesq\/ at the Fe K and \SNdsq\/ at the Nd L$_2$ edge, where \SFe\/ and \SNd\/ denote the magnetic order parameter for the Fe and the Nd sublattice, respectively.

The measured \SFesq\/ displays the typical behavior expected for a magnetic ordering transition and compares well to the corresponding neutron diffraction results (open circles in Fig. \ref{fig:Tdep152-fits}(e)). 

\SNdsq\/ exhibits a completely different temperature dependence and, in particular, does not show a clear phase transition at T$_N$. To understand the magnetic ordering behavior of the Nd sublattice, the corresponding curve in Fig.\ref{fig:Tdep152-fits}(f) was fitted to a mean field model, which assumes that the Nd order is merely induced by the Fe ordering\,\cite{PhysRevB.71.024414}:
\begin{eqnarray}
 |S_{\rm Nd}|^2 =
\left[S_{\rm Nd}^{\circ}~\tanh
\left(\frac{S_{\rm Nd}^{\circ}~\lambda_{\rm Fe-Nd}~3 \langle S_{\rm Fe}\rangle}{2k_BT}\right)\right]^2 
\label{eq:molecular-field}
\end{eqnarray} 
In this equation  $k_B$ is the Boltzmann constant, $\lambda_{\rm Fe-Nd}$ is the effective Fe-Nd coupling constant, $S_{\rm Nd}^{\circ}$ is the Nd ordered moment at $T$=0~K. Further, we used \SFesq$=\left( S^{\circ}_{\rm Fe}(1-T/T_N)^{\beta}\right)^2$ as a parameterization, where $S^{\circ}_{Fe}=5/2$ is the ordered Fe moment at $T=0$ and $\beta=0.31(2)$ was determined by a fit to the experimental data (red line in Fig.\ref{fig:Tdep152-fits}(e)).
The factor 3 accounts for the three Fe neighbors per Nd in the formula unit. Note that $\lambda_{\rm Fe-Nd}$\SFe~ is the magnetic mean field due to the Fe neighbors of the Nd sites that induces the magnetic order on the Nd sublattice.

The red solid line in Fig.\ref{fig:Tdep152-fits}(f) shows the fit of Eq.\,1 to the experimental data. The best fit is obtained using $S_{\rm Nd}^{\circ}=2.7\,\mu_B$ determined from neutron diffraction \cite{fischer06} and  $\lambda_{\rm Fe-Nd}=0.60(1)$~T/$\mu_B$. The excellent agreement with the data implies that the magnetic order on the Nd sublattice is induced by the driving Fe spin order.

\begin{figure}
\includegraphics[angle=-90,width=\columnwidth]{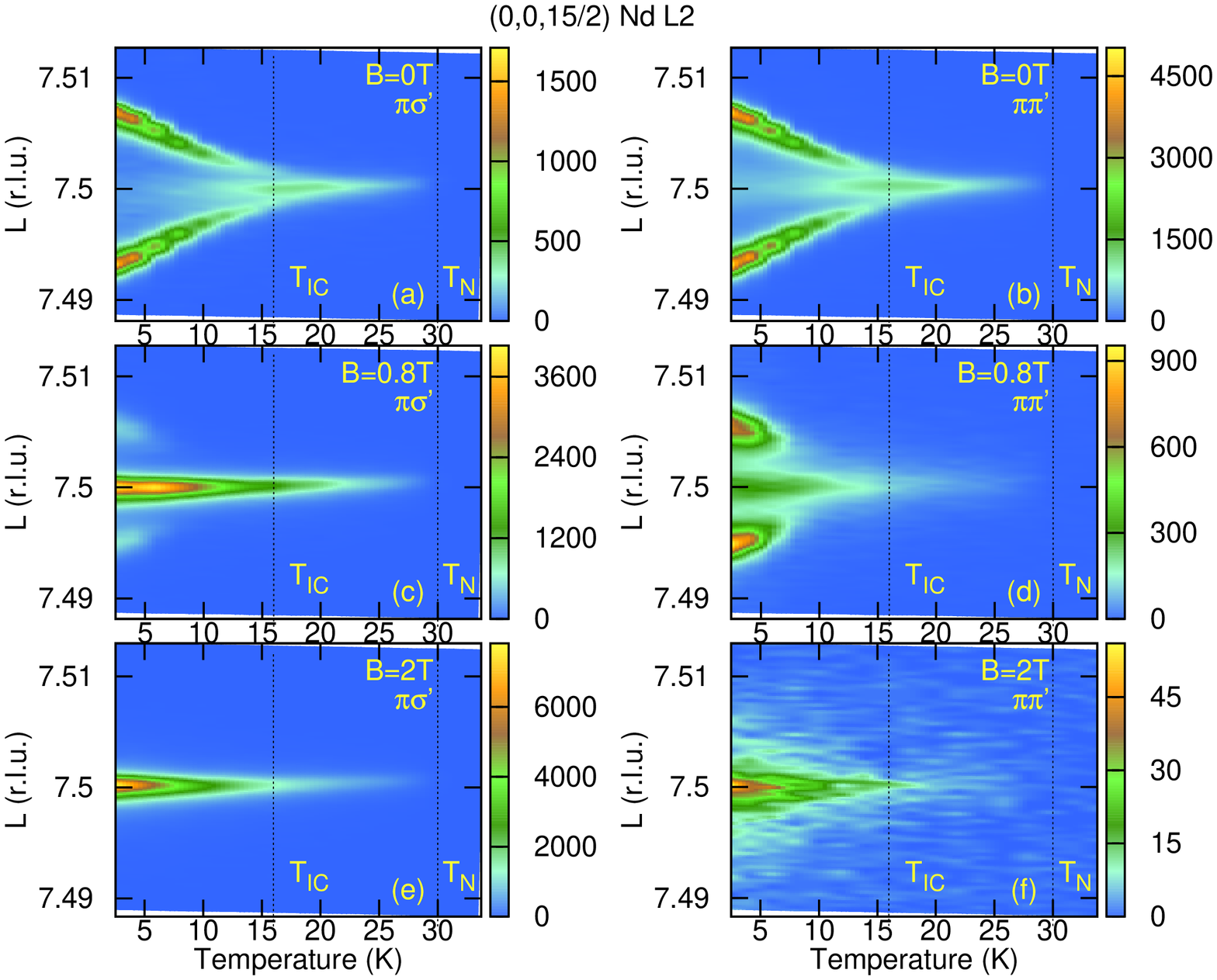}
\includegraphics[width=\columnwidth]{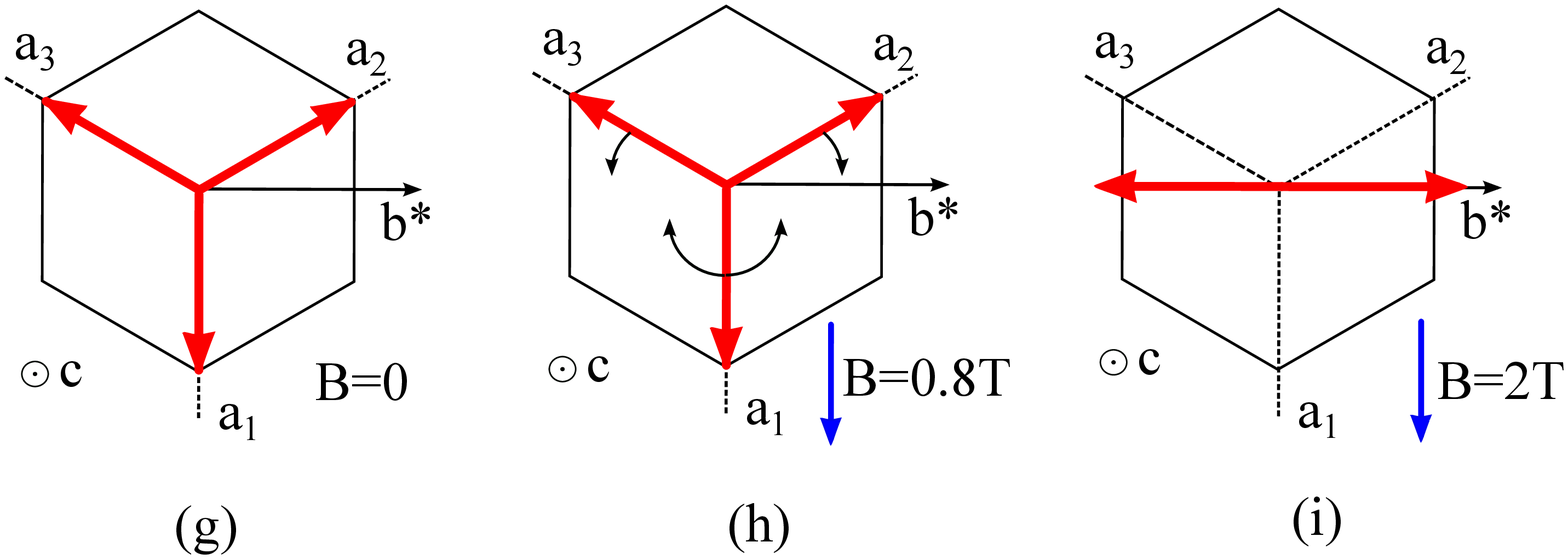} 
\caption{(Color online) Temperature dependence of the (0,0,15/2) peak in different applied magnetic fields. $\pi\sigma'$ and $\pi\pi'$ scattering is shown in the left and right panels, respectively. The magnetic field was always applied parallel to the $a$ axis. 
(g,h,i): Illustration of the spin directions as a function of magnetic field. Red arrows correspond to the easy axes of different magnetic domains in the $c$ phase.} 
\label{fig:15/2-Bdep}
\end{figure}

The intensity maps shown in Fig.\ref{fig:15/2-Bdep} display the temperature dependence of the magnetic (0,0,15/2) reflection measured  at the Nd L$_2$ edge in different applied magnetic fields. 
According to the previous neutron scattering studies, the magnetic moments of both Nd and Fe are always confined to the hexagonal $ab$ plane\,\cite{fischer06,janoschek}. 
In this situation, the point symmetry of the Nd sites implies that the RMXS of the Nd sites can be described within the formalism given in Ref.\,\onlinecite{Hannon1988,Hill},  i.e., the intensities $I_{\pi\pi'}$ and $I_{\pi\sigma'}$ can be expressed as: $I_{\pi\pi'}\propto(z_a\sin2\theta)^2$ and $I_{\pi\sigma'}\propto(-z_{b^*}\cos\theta)^2$, where $\theta$ is the scattering angle and $z_n$ are the projections of the magnetic moments along the crystallographic $a$ and $b^*$ directions (cf. Fig.\ref{fig:Tdep152-fits}(b)). 

As can be seen in Figs.\,\ref{fig:15/2-Bdep}a and \ref{fig:15/2-Bdep}b, no qualitative difference between the two intensities is observed for  \Bfield =0\,T apart from the fact that  $I_{\pi\pi'}/I_{\pi\sigma'}\sim3$. 
The relative intensities in the two scattering channels is the result of a particular domain structure. For instance, in the C phase the magnetic moments of the various domains are aligned along 3 equivalent directions within the hexagonal $ab$ plane (Fig.\ref{fig:15/2-Bdep}(g)). This yields $z_{a},z_{b^*}\neq0$ and intensities in both the $\pi\pi'$ and the $\pi\sigma'$ channel, where the value of  $I_{\pi\pi'}/I_{\pi\sigma'}$ depends on the volume fractions occupied by the different domains in the volume probed by x-rays. The magnetic domains will not be discussed in the following.
 
More importantly, there are already  remarkable changes in the RMXS for \Bfield =0.8\,T (Fig.\ref{fig:15/2-Bdep}(c) and \ref{fig:15/2-Bdep}(d)).
First, the IC phase is suppressed and only observed at lower temperatures. Second, there is an overall intensity transfer from $\pi\pi'$ to $\pi\sigma'$. This implies a reduction of  $z_a$ and an increase of $z_{b^*}$, i.e, the spins rotate as illustrated in Fig.\ref{fig:15/2-Bdep}(h). This observation agrees with magnetization\cite{hamann-boratesNRXS,volkov07-2} and sound velocity\cite{zvyagina_LTP2011} measurements which show that at \Bfield=0.8~T the system is in the center of a hysteresis corresponding to a spin flop transition. 

Increasing the magnetic field to \Bfield=2\,T,  the IC phase is completely suppressed, as can be seen in  Fig.\ref{fig:15/2-Bdep}(e) and \ref{fig:15/2-Bdep}(f). In this field, pure $\pi\sigma'$ scattering is observed,  whereas the $\pi\pi'$ is very weak, amounting only to 0.5\% of the $\pi\sigma'$ intensity. This implies that $z_a=0$, i.e., the spin moments are now essentially perpendicular to \Bvec ; i.e.; parallel to $b^*$ (Fig.\ref{fig:15/2-Bdep}(i)). Note also that the intensity in Fig.\ref{fig:15/2-Bdep}(e) is equal to the sum of the intensities measured in Fig.\ref{fig:15/2-Bdep}(a) and \ref{fig:15/2-Bdep}(b), showing that at \Bfield=2~T a single collinear magnetic domain is formed. 

\begin{figure}
 \includegraphics[width=0.4\columnwidth]{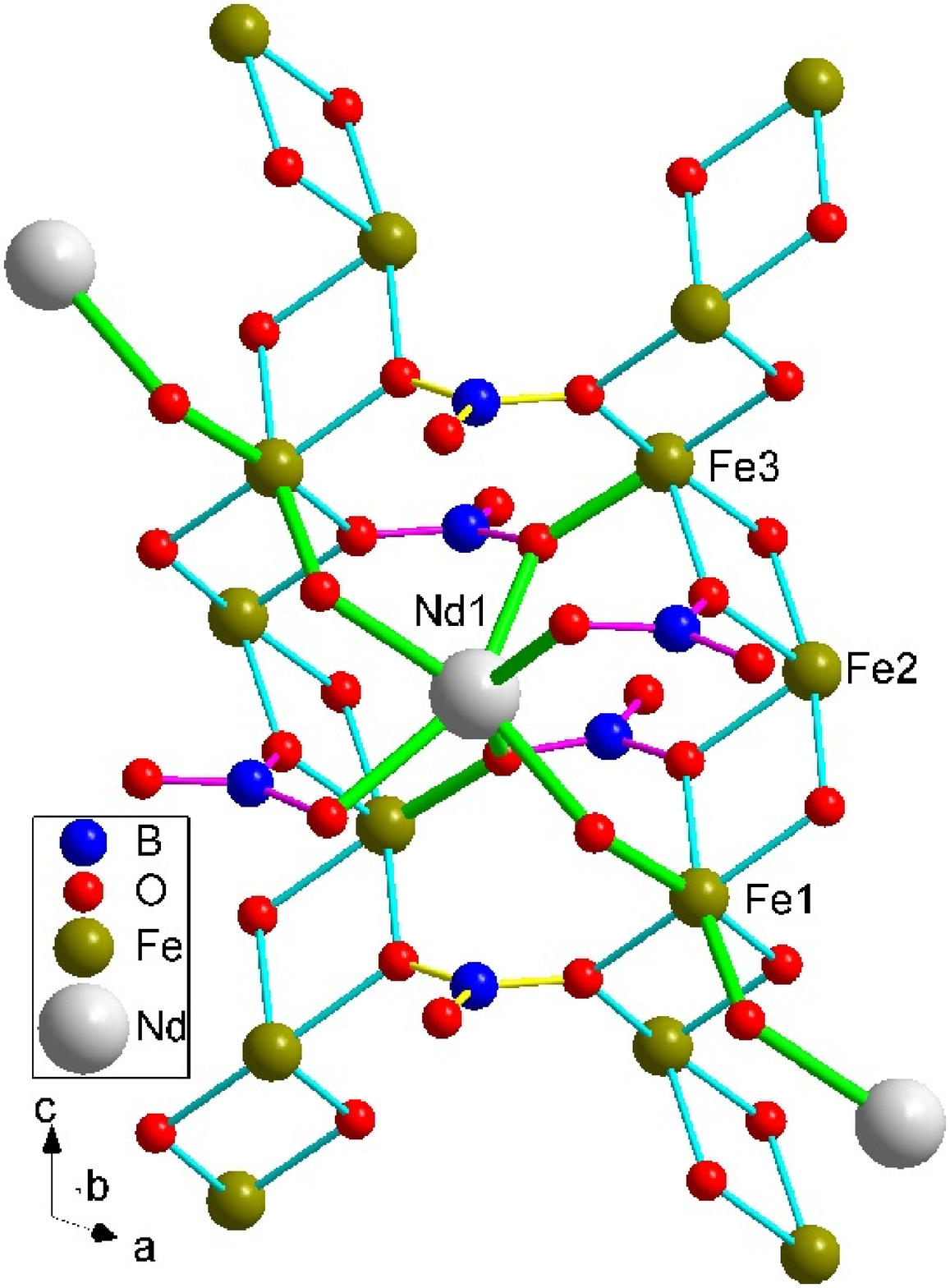}(a)
\includegraphics[width=0.4\columnwidth]{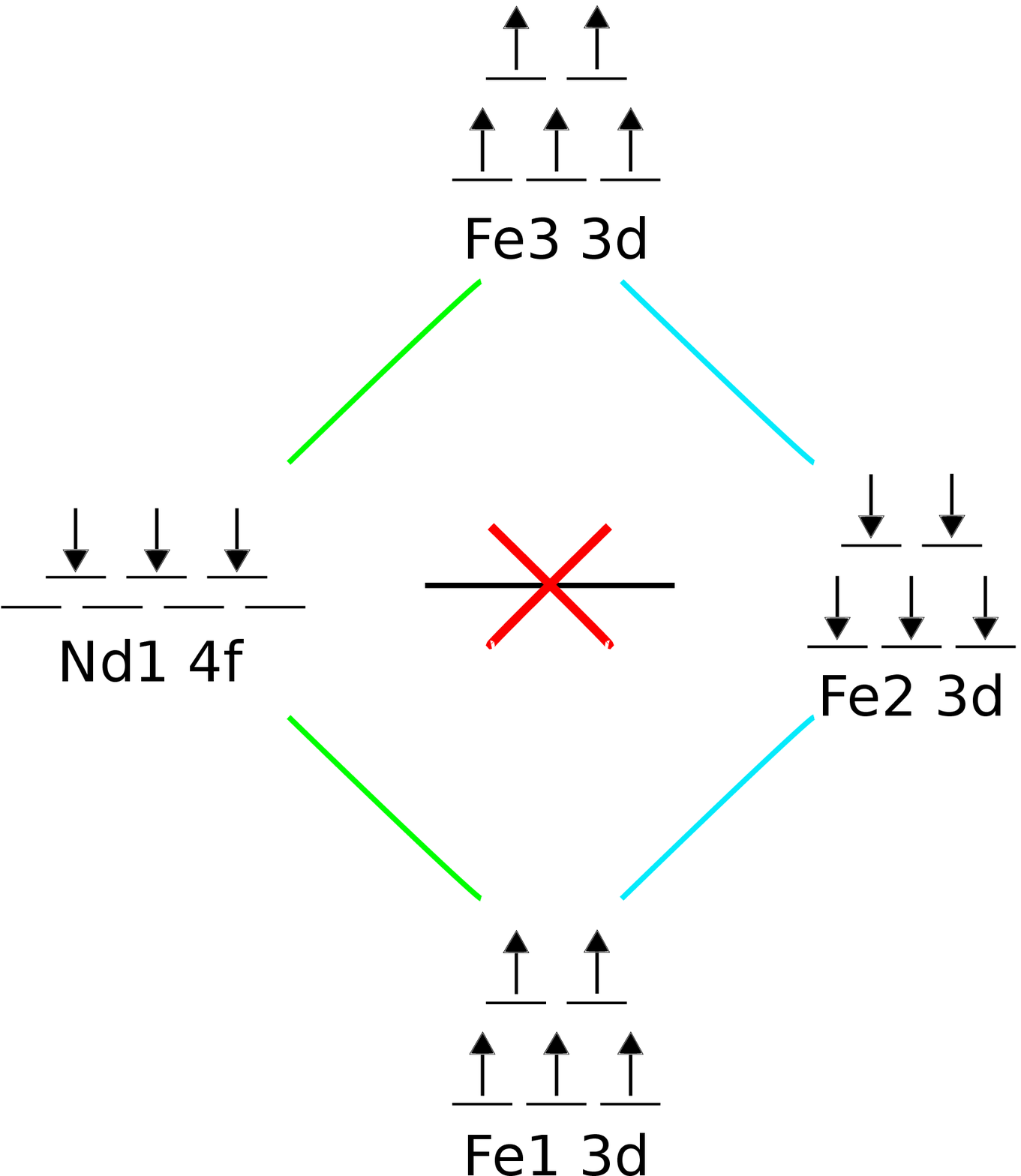}(b)
\caption{(Color online) (a) Crystal structure of \nd~ showing the next nearest neighbor superexchange paths between Fe-Fe (blue) and Nd-Fe (green). (b) Magnetic frustration between Nd1 and Fe2 ions. The strongest coupling is given by the intra chain Fe-Fe AFM coupling (blue). The Fe1,3-Nd1 (green) and Fe2-Nd1 (black) are expected to be weaker.} 
\label{fig:exchange-paths}
\end{figure}

These field dependent measurements reveal that the maximum of magnetic field induced ferroelectric polarization $\mathbf{P}$ coexists with collinear spin order. The spin helix can hence be excluded as an origin for the magneto-electric effect in \nd . 

The microscopic insights obtained by RMXS enable to deduce a consistent scenario for the magneto-electric coupling in \nd , which is based on the magnetic frustration between the Fe and Nd sublattices: according to previous reports\,\cite{fischer06}, Fe has a $3d^5$ configuration, which implies that both the Fe-Fe and the Nd-Fe superexchange interactions are antiferromagnetic (AFM). Our data further shows that \SFesq\/ corresponds to the primary order parameter, whereas \SNdsq\/ is merely induced. We therefore consider the coupling of Nd to fully ordered Fe chains. As illustrated in Fig.\,\ref{fig:exchange-paths}, the AFM Nd-Fe interactions are frustrated, even though not all Nd-Fe couplings have the same strength. Upon decreasing temperature \SNdsq\/ increases, i.e., an average magnetic moment on the Nd sites develops. This finite  \SNdsq\/  then also increases the energy related to the Nd-Fe frustration in the collinear phase.  

Without applied magnetic field, this Nd-Fe frustration can be relieved by forming a spin helix, which optimizes the magnetic interactions and gains exchange energy $\Delta E_{IC}$ by slight rotations $\Delta \phi$ of neighboring spins along $c$.  Since \SNdsq\/ increases upon cooling, $\Delta E_{IC}/  \Delta \phi$ will also depend on temperature, which naturally explains the strongly temperature dependent period $\lambda\propto 1/\Delta\phi$ of the spin helix. We therefore conclude that the spin helix is driven by the Nd-Fe sublattice frustration. 

By applying a magnetic field parallel to the $ab$-plane, the spin helix is rapidly suppressed and the spin rotations are no longer energetically favorable. We propose that within the magnetic field enforced C phase, the system choses an alternative way to relieve the magnetic frustration, which involves a modulation of the lattice structure and creates a FE polarization. We cannot determine these distortions based on the present data, but the 3-fold rotation axis of the space group $R32$ has to be broken in order to allow for a ferroelectric polarization parallel to the $ab$-plane. Interestingly, this removes the symmetry relations exactly between the Fe-sites shown in Fig. \ref{fig:exchange-paths}(a)~\footnote{The maximal non isomorphic subgroup that allows such symmetry reduction is $C2$, corresponding to a monoclinic setting. For instance, Fe1, Fe2 and Fe3 in Fig.\ref{fig:exchange-paths}(a) are symmetry related in the $R32$ space group but they are not in $C2$. Thus atomic displacements of these ions (or ligand ions attached to them) might induce some electric polarization parallel to the $ab$-plane.}. The required symmetry reduction is therefore fully consistent with the deduced microscopic mechanism, which further supports our conclusions.
 
\nd\/ hence belongs to a class of materials where the magneto-electric properties are driven by the magnetic frustration of two interacting subsystems and where the degree of frustration can even be tuned by changing temperature. This frustration can be relieved either by forming a spin helix or by creating a FE polarization. The FE collinear and the non-FE spin helix are therefore competing phases. Due to this competition an external magnetic field can easily switch between the two phases, causing the ME effect. We believe that a quantitative theoretical analysis of this mechanism may help to identify new magnetoelecric materials with optimized properties. The presented RMXS study also beautifully illustrates the power of this experimental technique and shows how this enables to unveil hidden microscopic physics that can hardly be accessed otherwise. This strongly motivates further studies of materials with different interacting magnetic sublattices using RMXS.

The authors would like to thank C. Detlefs and J. Trinckauf for fruitful discussions.  For the technical support during experiments we would like to thank H. Walker and the ESRF and BESSY for beamtime provision. This work was supported by the DFG through the Emmy Noether Programme (Grant GE1647/2-1) and the European Community's Seventh Framework Program (FP7/2007-2013) under grant Nr. 226716. 


%

\end{document}